\documentclass[conference,11pt]{IEEEtran}
\usepackage[pass]{geometry}  % 'pass' prevents scaling text
\IEEEoverridecommandlockouts

\usepackage{xcolor}
\definecolor{boriscomments}{rgb}{0.88, 0.04, 0.04}
\definecolor{ferrancomments}{rgb}{0, 0.5, 0.8}

\usepackage[acronym,toc,nomain]{glossaries} 
\makeglossaries
\newacronym{ap}{AP}{Access Point}
\newacronym{ai}{AI}{Artificial Intelligence}
\newacronym{xr}{XR}{eXtended Reality}
\newacronym{vr}{VR}{Virtual Reality}
\newacronym{ar}{AR}{Augmented Reality}
\newacronym{vf}{VF}{Video Frame}
\newacronym{hmd}{HMD}{Head Mounted Display}
\newacronym{alvr}{ALVR}{Air Light VR}
\newacronym{qos}{QoS}{Quality of Service}
\newacronym{sinr}{SINR}{Signal-to-interference-plus-Noise ratio}
\newacronym{snr}{SNR}{Signal plus Noise Ratio}
\newacronym{per}{PER}{Packet Error Rate}
\newacronym{mcs}{MCS}{Modulation and Coding Scheme}
\newacronym{dcf}{DCF}{Distributed Coordination Function}
\newacronym{sifs}{SIFS}{Short Interframe Space}
\newacronym{difs}{DIFS}{DCF interframe space}
\newacronym{cw}{CW}{Contention Window}
\newacronym{mpdu}{MPDU}{MAC Protocol Data Unit}
\newacronym{ampdu}{A-MPDU}{Aggregate MAC protocol Data Unit}
\newacronym{back}{BACK}{Block ACKnowledgment}
\newacronym{sta}{STA}{station}
\newacronym{bss}{BSS}{Basic Service Set}
\newacronym{ofdm}{OFDM}{Orthogonal Frequency-Division Multiplexing}
\newacronym{ofdma}{OFDMA}{Orthogonal Frequency-Division Multiple Access}
\newacronym{fifo}{FIFO}{First In First Out}
\newacronym{edca}{EDCA}{Enhanced Distributed Channel Access}
\newacronym{nav}{NAV}{Network Allocation Vector}
\newacronym{ul}{UL}{Uplink}
\newacronym{dl}{DL}{Downlink}
\newacronym{cu}{CU}{Channel Utilization}
\newacronym{flr}{FLR}{Frame Loss Ratio}
\newacronym{gop}{GoP}{Group of Pictures}
\newacronym{qoe}{QoE}{Quality of Experience}
\newacronym{mtp}{MTP}{Motion-to-Photon}
\newacronym{rtt}{RTT}{Round Trip Time}
\newacronym{vfrtt}{VF-RTT}{Video Frame Round Trip Time}
\newacronym{rts}{RTS}{Request-to-Send}
\newacronym{cts}{CTS}{Clear-to-Send}
\newacronym{mimo}{MIMO}{Multiple-Input Multiple-Output}
\newacronym{nal}{NAL}{Network Abstraction Layer}
\newacronym{abr}{ABR}{Adaptive Bitrate}
\newacronym{udp}{UDP}{User Datagram Protocol}
\newacronym{rssi}{RSSI}{Received Signal Strength Information}
\newacronym{csmaca}{CSMA/CA}{Carrier Sense Multiple Access with Collision Avoidance}
\newacronym{cbr}{CBR}{Constant Bitrate }
\newacronym{ir}{IR}{Intra-refresh}
\usepackage[T1]{fontenc}
\usepackage{url}
\usepackage{soul}
\usepackage{amsmath,amssymb,amsfonts}
\usepackage{newtxtext,newtxmath}
\usepackage{textcomp}
\usepackage{siunitx}
\usepackage{float}
\usepackage{graphicx}
\usepackage{subcaption}
\usepackage{cite}
\usepackage[T1]{fontenc}
\usepackage{lmodern} % Latin Modern: scalable Type1 fonts for fallback/sans/serif

\usepackage{booktabs}
\usepackage{multirow}
\usepackage{tablefootnote}

\usepackage[linesnumbered,ruled,vlined]{algorithm2e}
\SetKwComment{Comment}{\{}{\}}

\usepackage[numbers]{natbib}
\begin{document}

% This code is to reduce the list of authors by using et. al:
\bstctlcite{IEEEexample:BSTcontrol}

%\title{ Modelling bursty HEVC traffic over Wi-Fi in Multi-user Virtual Reality Streaming Applications\
%\title{How many VR Streams can Wi-Fi handle? \\A Study with Realistic Video Traffic\
% \title{Performance Evaluation of VR Real-Time Streaming in Wi-Fi: An HEVC Analysis \
% \title{Pushing the Limits of Wi-Fi for Multi-user VR Streaming: An HEVC-based Analysis \
% \title{Evaluating HEVC Coding Strategies for Multi-user VR Streaming Over Wi-Fi Networks \
\title{Evaluating Wi-Fi Performance for VR Streaming: \\ A Study on Realistic HEVC Video Traffic
 % \thanks{
 % This work is partially funded by Wi-XR PID2021-123995NB-I00 (MCIU/AEI/FEDER,UE), by MCIN/AEI under the Maria de Maeztu Units of Excellence Programme (CEX2021-001195-M), and by AGAUR ICREA Academia 00077, and by an FPI fellowship (PRE2022-101847) funded by MCIN/AEI/
 % 10.13039/501100011033. }}
\thanks{
This work is partially funded by Wi-XR PID2021-123995NB-I00 and TRUE-Wi-Fi PID2024-155470NB-I00 (MCIU/AEI/FEDER,UE) (MCIU/AEI/FEDER, UE), by MCIN/AEI under the Maria de Maeztu Units of Excellence Programme (CEX2021-001195-M), and by the Spanish Ministry of Science, Innovation and Universities under an FPI fellowship (PRE2022-101847).
}}

 \author{\IEEEauthorblockN{Ferran Maura, Francesc Wilhelmi, Boris Bellalta}
 \IEEEauthorblockA{Universitat Pompeu Fabra, Barcelona, Spain
 \textit{Wireless Networking Research Group, Universitat Pompeu Fabra}\\
    Carrer de Roc Boronat 138, 08018 Barcelona, Spain
 \\
 \{ferranjosep.maura, francisco.wilhelmi, boris.bellalta\}@upf.edu}
 }
\maketitle
\begin{abstract}
Cloud-based Virtual Reality (VR) streaming presents significant challenges for 802.11 networks due to its high throughput and low latency requirements. When multiple VR users share a Wi-Fi network, the resulting uplink and downlink traffic can quickly saturate the channel. This paper investigates the capacity of 802.11 networks for supporting realistic VR streaming workloads across varying frame rates, bitrates, codec settings, and numbers of users. We develop an emulation framework that reproduces Air Light VR (ALVR) operation, where real HEVC video traffic is fed into an 802.11 simulation model. Our findings explore Wi-Fi's performance anomaly and demonstrate that \gls{ir} coding effectively reduces latency variability and improves QoS, supporting up to 4 concurrent VR users with \gls{cbr} 100 Mbps before the channel is saturated.
\end{abstract} 


% \maketitle

% \tableofcontents 

% ------------------------------
% ------------------------------
% ------------------------------
% ------------------------------
\section{Introduction}

\Gls{xr}---which encompasses both \gls{vr} and \gls{ar}---is expected to be a transformative technology in the coming years. While \gls{xr} is already well-established in gaming, it is poised to disrupt a wide range of emerging use cases, including immersive remote collaboration, virtual training and education, remote medical procedures, and teleoperation of industrial machinery~\cite{alriksson2021xr}.

Cloud \gls{vr} streaming is a bandwidth-intensive \gls{xr} application in which interactive VR content is rendered on powerful remote servers (e.g., in the cloud) and streamed to less capable end devices. These servers must provide sufficient CPU and GPU resources to render scenes at high frame rates and run the \gls{vr} streaming application in real time. Specifically, the server encodes and transmits video and audio streams---at their respective frame and sampling rates---to the user's \gls{hmd}. Simultaneously, the \gls{hmd} tracks the user's pose and movements, sending tracking data back to the server to update the player's state. Based on this feedback, new frames are rendered at the server and streamed back to the \gls{hmd} for display~\cite{ALVR_HowALVRWorks}.

Cloud rendering offers the advantage of reducing client-side hardware requirements. As long as the end-to-end latency remains sufficiently low, \gls{vr} streaming enables high-fidelity graphics that would otherwise exceed the rendering capabilities of typical \gls{hmd} devices. While wired connections can support such applications, wireless technologies---particularly IEEE 802.11 (commercially known as Wi-Fi) 
% ~\cite{geraci2025wi}
, the most widely used standard for indoor connectivity---are increasingly preferred for their support of user mobility and comfort. However, meeting the strict latency and reliability requirements of real-time \gls{vr} streaming is challenging for wireless networks. Wi-Fi, in particular, faces several limitations. First, 802.11 implements random channel access through the \gls{dcf}, meaning that devices must compete for access to the medium. Second, the server, \gls{ap} and \gls{hmd} \glspl{sta} do not coordinate their \gls{dl} and \gls{ul} transmissions, which can further degrade the quality of the \gls{vr} experience~\cite{carrascosa2024performance}. Finally, interference from neighboring networks operating on the same channel can introduce unpredictable disruptions, further reducing the performance of \gls{vr} applications~\cite{casasnovas2025can}.
\newline
In this paper, we study the suitability of running \gls{vr} streaming applications over Wi-Fi. The main novelty of this study lies in the use of real encoded video transmissions within 802.11 simulations, which is key to accurately reproducing frame sizes resulting from different codec configurations. Our main contributions are the following:
\begin{itemize}
    \item \textbf{\gls{vr} streaming emulation over IEEE 802.11:}  
    We develop a modular simulation framework that emulates one or multiple \gls{vr} streaming sessions between a server and an \gls{hmd} client over a shared Wi-Fi network. The framework includes encapsulated applications for the server and client, a video codec pipeline capable of processing 4K video at configurable frame rates and bitrates, and a queueing and transmission model representing IEEE 802.11 PHY/MAC behavior.
    \item \textbf{Joint analysis of network and codec-level dynamics:}  
    We use our integrated simulation framework to evaluate how key \gls{qos} metrics (e.g., latency, packet loss) are affected by codec parameters (e.g., frame rate, bitrate), network contention (e.g., number of \gls{cbr} users), and PHY conditions (e.g., \gls{sta}-to-AP distance). Our results offer insights into the interplay between video encoding strategies and wireless performance in dense \gls{vr} scenarios.
\end{itemize}

% ------------------------------
% ------------------------------
% ------------------------------
% ------------------------------
\section{Cloud-based VR streaming}

In this section, we introduce the architecture of cloud \gls{vr} streaming, provide an overview of the video coding strategies currently used, and discuss the performance requirements for ensuring a good user experience. 

% ------------------------------
% ------------------------------
\subsection{Server-Client Architecture}

In \gls{vr} streaming, rendering typically takes place on a server, which encodes and transmits video, audio, and haptic data to an \gls{hmd}. The \gls{hmd} continuously tracks the user's pose and movements, sending tracking packets to the server. This feedback enables the server to update the user's position, render the scene accordingly, and transmit the next frames to the \gls{hmd} for display.

Representative examples of this architecture include \gls{alvr}\footnote{\url{https://github.com/alvr-org/ALVR}} and WiVRn\footnote{\url{https://github.com/WiVRn/WiVRn}}. \gls{alvr} is an MIT-licensed application for \gls{vr} streaming that acts as a wireless bridge between a SteamVR server and an \gls{hmd} client. The server transmits video, audio, and haptic data, while the client sends tracking information and statistics back to the server. \gls{alvr} also supports \gls{vr}-specific features such as foveated rendering—reducing resolution on image borders for more efficient encoding—and reprojection, which helps compensate for rapid head movements not yet rendered by the server~\cite{ALVR_HowALVRWorks}.

% ------------------------------
% ------------------------------
\subsection{Video Coding for Real-Time Streaming}

Video traffic between server and client is typically compressed, drastically reducing bandwidth requirements compared to raw RGB video. However, compression introduces encoding and decoding latencies for each \gls{vf}, which contribute to increasing the end-to-end latency. Modern codecs such as H.264 and HEVC typically encode frames into a \gls{gop}, consisting of intra-coded frames (I-frames) and predicted frames (P-frames)~\cite{richardson2024coding}. For real-time streaming, B-frames (bidirectionally predicted frames) are not considered, as they require access to future frames at the encoder side. 

% For instance, a \gls{gop} size of 10 implies that one I-frame is followed by nine P-frames.

% I-frames encode full image data and serve as reference points, while P-frames encode only the differences from preceding frames. Losing a P-frame introduces visual artifacts from error propagation, and losing an I-frame can prevent decoding of subsequent P-frames until a new I-frame is received. These artifacts typically disappear once a new I-frame is correctly decoded. I-frames can be inserted automatically by an encoder when a scene change is detected, or requested by a client to recover from frame loss. 

HEVC’s \gls{ir} mode offers improved resilience by gradually injecting I-frame coding units into successive frames. This not only enhances robustness under frame loss conditions~\cite{CHEN2015294, Intrarefresh_BER_HEVC} but also results in a less variable frame size distribution~\cite{HowtoModelVR}.

% ------------------------------
% ------------------------------
\subsection{Requirements for High QoE}

Due to the proximity of the \gls{hmd} screens to the user’s eyes, \gls{vr} streaming requires high resolutions and bitrates to ensure visual fidelity for a good \gls{qoe}. The \textit{bitrate} specifies the compression level applied during encoding, and along with the frame rate (FPS), determines the average frame size and visual quality. In this study, we consider bitrates ranging from 10 to 100 Mbps.

Regarding latency, we distinguish between: a) \gls{mtp} latency—i.e., the delay between a physical user movement and its visual update in the \gls{hmd}; and  b) network latency. For (a), ITU-T recommends that \gls{mtp} latency should remain below 50~ms for 360º \gls{vr} experiences~\cite{ITU-TJ.1631-2021}. This constraint includes time for rendering, encoding, network transport and decoding. For (b), use cases involving strong interaction, such as cloud \gls{vr} gaming, require a network \gls{rtt} $\leq 20$~ms. 

For ALVR, the effect of \gls{dl} video packet loss is such that if any packet that carries a fragment of a \gls{vf} is lost, the entire frame is considered lost. With typical bitrates, each \gls{vf} may consist of dozens of packets. Hence, even a seemingly small constant loss probability of 0.01\% can result in a \gls{flr} of around 1\%. 

According to~\cite{casasnovas2025nestvradaptivebitratealgorithm}, for a seamless ALVR user experience, the \gls{vfrtt} should not exceed 33~ms, and at least 99\% of transmitted frames should be successfully received. In our evaluation we adopt a \gls{flr} $\leq 1\%$ and network latency $\leq 33$~ms as thresholds for acceptable \gls{qoe}.  
\glsreset{flr}

% ------------------------------
% ------------------------------
% ------------------------------
% ------------------------------
\section{Emulation Framework}

\begin{figure*}[!t] % single-column float, try to place at top
  \centering
  \includegraphics[width=0.8\textwidth]{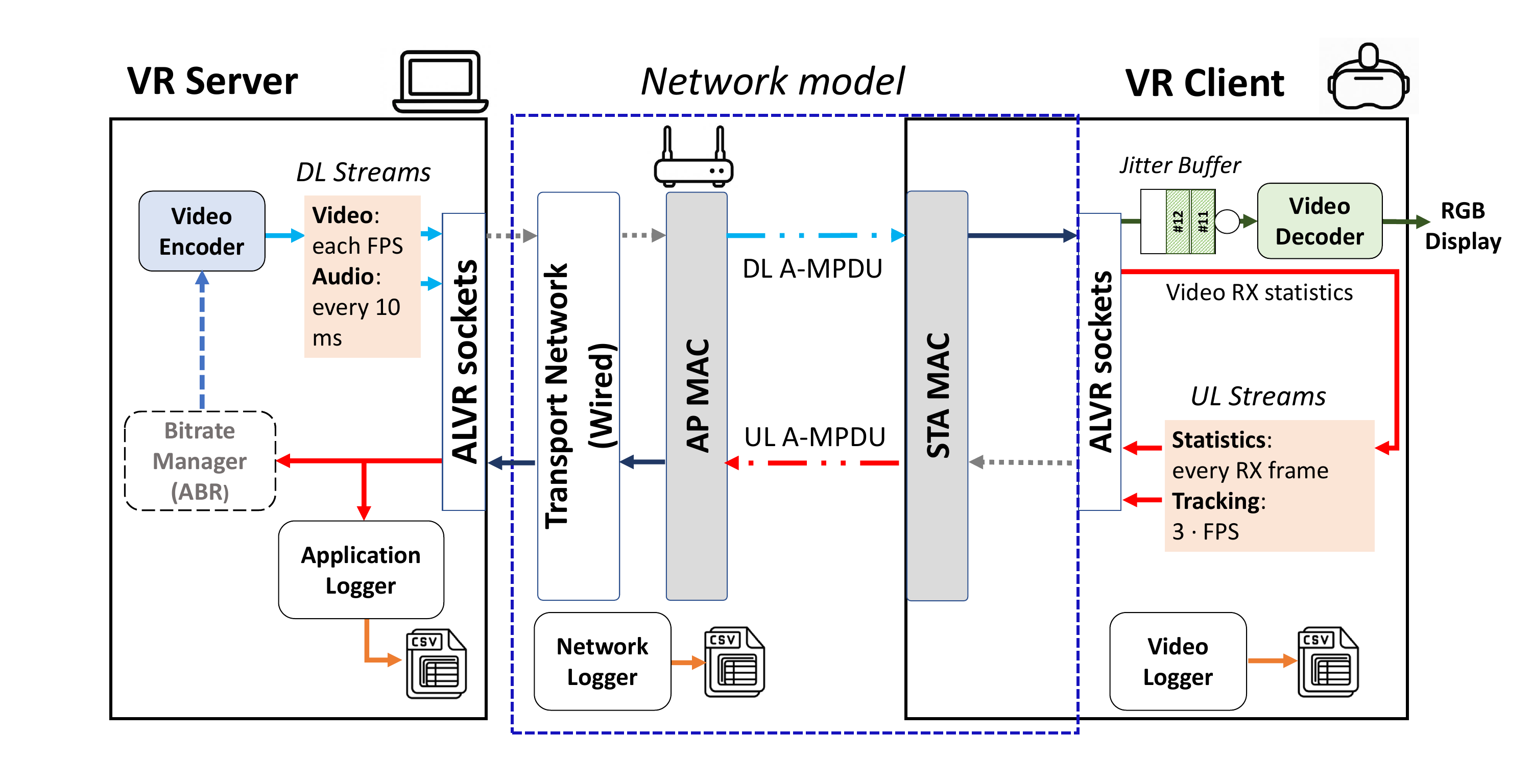} % renamed to avoid spaces
  \caption{Diagram of our proposed framework for emulating multiple \gls{vr} streaming sessions.}
  \label{fig:Framework_diagram}
\end{figure*}

In Fig.~\ref{fig:Framework_diagram}, we illustrate the functionality of each main component (i.e., server, client, \glspl{sta}, and \gls{ap}) in the proposed framework. The framework is implemented in Rust using the NexoSim library~\cite{asynchronicsnexosim2025} for discrete-event simulation.

Simulation inputs (e.g., number of users, device positions, bitrate, FPS) are configured via a Bash script, which also automates scenario iteration by combining different input parameters. For each emulated streaming session, a pair of server and client \texttt{Model} instances~\cite{asynchronicsnexosim2025} is initialized, with their respective addresses and connections to \glspl{sta} and \gls{ap}. These server-client blocks are decoupled from the network model, which separately characterizes the behaviour of 802.11 devices on a shared channel. Logging modules, embedded in the \gls{ap} and server/client blocks, store per-user metrics in CSV files for each scenario. The Bitrate~Manager block at the server is reserved for future work; \gls{cbr} is assumed.

The behavior of \gls{alvr} is emulated by using its native data structures and real application-level \gls{dl}/\gls{ul} transmission rates. All application streams, except control traffic, are exchanged between the server and client using instances of \gls{alvr} \textit{stream sockets}. Before sending data through the transport protocol, each payload is encapsulated in an \gls{alvr}-compliant structure. This includes packet fragmentation and tagging with an application-layer header that carries metadata such as stream type and sequence number. On the receiving side, stream sockets reconstruct the original frames from received fragments. This native structure is preserved to enable realistic metrics computation at the \gls{hmd} application level.

Each simulated scenario starts with server and client instances paired and ready to initiate streaming. Streaming sessions are triggered by shared events between simulation models. Details of the server and client implementations follow.

%The behavior of \gls{alvr} in the network is emulated by using its native data structures in traffic, and using \gls{dl}/\gls{ul} transmission rates of the real application as a basis. 
%All streams except control traffic are shared between the server/client block using instances of \gls{alvr} \textit{stream sockets}. Before sending data through a transmission protocol of choice, an \gls{alvr}-compliant structure is adopted to fragment and tag packets with an application header, which contains information about the stream type, their packet sequence number, and other information to allow reconstructing the original fragmented data. On the receiver side, \gls{alvr} sockets receive the stream of data and reconstruct fragmented frames. This native structure is preserved in the framework for the sake of computing realistic metrics on each \gls{hmd} application.

% ------------------------------
\subsection{VR Server}

The roles of SteamVR and \gls{alvr}'s split rendering are replaced by a configurable encode/decode pipeline, capable of handling 4K video at variable frame rates, bitrates, and transcoding parameters. Each \gls{dl}/\gls{ul} stream in a session uses a dedicated stream socket, allowing video and audio to be independently scheduled on the server.

Audio packets are transmitted every 10~ms in pairs via the audio stream socket, each with empty content and a combined size of 2000~bytes. Video is generated by invoking FFmpeg~\cite{ffmpeg2025} to transcode sample content into encoded chunks of $T_{\text{CHUNK}}$ seconds, with user-defined settings (e.g., codec preset, bitrate, GoP size, \gls{ir}). The encoded output is stored in a video buffer.

Single HEVC \gls{nal} units are extracted from this buffer at a constant FPS, fragmented into packets, and sent in bursts via the stream socket. When the buffer is exhausted, a new chunk is transcoded, and the cycle repeats. Chunk-based encoding allows for easier integration of \gls{abr} algorithms and reduces computational overhead compared to frame-by-frame encoding. Additional details on the video transcoding process are provided in Section~\ref{sec:system_model}.

% ------------------------------
\subsection{VR Client}

The client periodically generates and transmits single \gls{ul} tracking packets at a rate of \(3 \times \text{FPS}\). Upon receiving all fragments of a video frame, the \gls{alvr} receive socket reconstructs the original encoded \gls{vf}. If a frame is not fully reconstructed within 0.1 seconds of receiving its first fragment, it is discarded, and a frame loss report is immediately sent via an \gls{ul} packet.

Successfully reconstructed frames contribute to the next UL statistics packet, and encoded \glspl{vf} are placed into a \textit{jitter buffer} of capacity $B_\text{DEC}$. This buffer temporarily stores frames prior to display to mitigate network delay variations and maintain a consistent playback rate. A simulated \textit{vsync} mechanism retrieves frames from the buffer at a constant FPS, decodes them, and presents the final RGB output.

% ------------------------------
% \subsection{Transport Network}

% The transport network block connects each \gls{vr} server to the \gls{ap} and serves as the interface for introducing network effects based on \texttt{tc-netem}.\footnote{\url{https://man7.org/linux/man-pages/man8/tc.8.html}} It supports emulation of packet loss, jitter, and bandwidth constraints with configurable timing and probability. No transport-layer impairments are introduced in this paper; the feature is reserved for future experiments.

% ------------------------------
\subsection{Wi-Fi Network}

Wi-Fi is emulated via a discrete-event simulator designed to represent IEEE 802.11be (Wi-Fi~7) networks.\footnote{While advanced features such as Multi-Link Operation (MLO) are not yet included, all base IEEE 802.11be parameters are used in this study.} The model includes \gls{ap} and \gls{sta} operations and implements the \gls{dcf} with an RTS/CTS mechanism. It supports multiple transmission rates determined by received power according to the 802.11be \gls{mcs} tables, as well as retransmissions, single-user \gls{mimo}, and packet aggregation via \glspl{ampdu}.

% ----------------------------------------
% ----------------------------------------
% ----------------------------------------
% ----------------------------------------
\section{System Model}
\label{sec:system_model}

% ------------------------------
% ------------------------------
\subsection{VR Emulation Framework Setup}

We consider a Wi-Fi network consisting of one \gls{ap} and $N$ \glspl{sta}, each placed at a configurable distance from the \gls{ap}. As in a typical \gls{vr} streaming setup, each \gls{sta} runs a client application paired with a dedicated server, which is connected to the \gls{ap} via a wired link. All traffic flows are carried over \gls{udp}.

To compute received power for each transmission, we use the TMB indoor path loss model from~\cite{TMBPathloss}, which is then used to select the \gls{mcs} per user. The highest achievable rate is $\text{MCS}_{13}$, selected when \gls{rssi} exceeds $-46$~dBm. We assume single-user transmissions and do not include features such as \gls{ofdma}. Transmission delays account for the temporal overhead introduced by \gls{csmaca} with a \gls{rts}/\gls{cts} scheme for channel access. Each data packet within an \gls{ampdu} has a 10\% probability of error, leading---along with collisions---to MAC-layer retransmissions.

To generate streaming content, we use two 3840$\times$2160 resolution video samples at 60~FPS: \textit{BigBuckBunny}, a standard 3D animated short in stereo video,\footnote{\url{http://distribution.bbb3d.renderfarming.net/video/mp4/bbb_sunflower_2160p_60fps_stereo_abl.mp4}} and \textit{snow}, a high-motion POV video of professional snowboarding scenes. Their average bitrates, measured using \texttt{ffprobe}, are $12.006$~Mbps and $86.627$~Mbps, respectively.

To emulate higher frame rates without requiring high-FPS video samples, we scale the bitrate used in the FFmpeg encoding command proportionally to the ratio between the simulated FPS and the original sample FPS. Although FFmpeg has support for other modern codecs (e.g., AV1, H.264), for consistency and brevity, HEVC with the \textit{fast} preset is used for all experiments.  

% ------------------------------
% ------------------------------
\subsection{Performance Metrics Considered}

To evaluate the performance of \gls{vr} applications under various conditions, we consider the following metrics:
\begin{itemize}
    \item \textbf{Video Frame Round-Trip Time (VF-RTT):} The time duration for the server to transmit a complete \gls{vf}, plus the time until the corresponding \gls{ul} statistics packet is received at the server~\cite{casasnovas2025nestvradaptivebitratealgorithm}. This metric captures the application's experienced network latency.
    \item \textbf{\gls{cu}:} Computed as the ratio between the time duration spent on wireless transmissions and collisions, and the overall simulation duration.
    \item \textbf{\gls{flr}:} Defined as the proportion of \glspl{vf} that were not fully received by the client, relative to the total number of frames transmitted by the server.
    \item \textbf{VMAF Score:} Based on the Netflix Video Multi-Method Assessment Fusion model (VMAF)~\cite{NetflixVMAF}, this score is computed by comparing each decoded RGB frame in the \gls{hmd} with a high-bitrate reference (100~Mbps) of the same video. Both versions share the same \gls{ir} or \gls{gop} configuration. The resulting perceptual quality scores range from 0 to 100.
\end{itemize}

% ------------------------------
% ------------------------------
% ------------------------------
% ------------------------------

\section{Results}

This section leverages the proposed framework to evaluate Wi-Fi's ability to support timely \gls{vr} traffic. By varying network and codec parameters, and monitoring application and network metrics, we assess the impact of different \gls{vr} session configurations, number of users, and heterogeneous \gls{mcs} rates. Key simulation parameters are listed in Table~\ref{tab:simulation-params}.

\begin{table}[!ht]
\centering
\caption{Simulation parameters.}
\label{tab:simulation-params}
\begin{tabular}{ll}
\hline
\textbf{Parameter} & \textbf{Value} \\ \hline
Simulation time & $20$~s \\ 
% \gls{ap} Queue size & $10000$~packets \\
Frequency band & $5$~GHz \\
Channel width & $80$~MHz \\
Transmit power & $20$~dBm \\
Maximum A-MPDU size & $64$~packets \\
Maximum A-MPDU transmission time & $4.85$~ms \\
Contention Window & $8$~slots \\
%Legacy PHY duration & $20\ \mu$s 
%PHY duration & $100\ \mu$s \\
%Slot time & $9\ \mu$s \\
\hline
FFmpeg version & 7.1 \\
Video chunk duration ($T_{\text{CHUNK}}$)     & $1.5$~s \\
% Video codec                      & HEVC (libx265, nvenc) \\
% GoP structure & I-P or \gls{ir}\\
Jitter buffer capacity ($B_{\text{DEC}}$) & $2 $\,frames  \\
VMAF model & \texttt{libvmaf} 4K v0.6.1 \\
\hline
\end{tabular}
\end{table}

\begin{figure}[!h]
    \centering
  \includegraphics[width=0.99\linewidth]{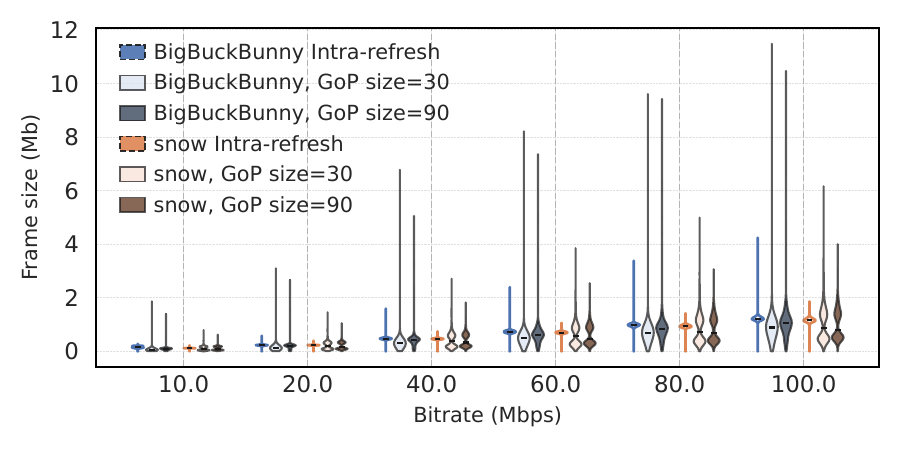}
    \includegraphics[width=0.99\linewidth]{
    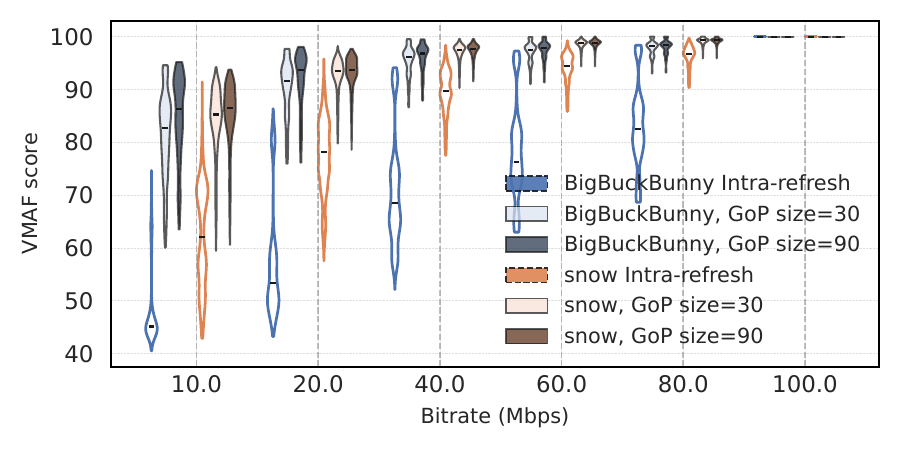}
    \includegraphics[width=0.99\linewidth]{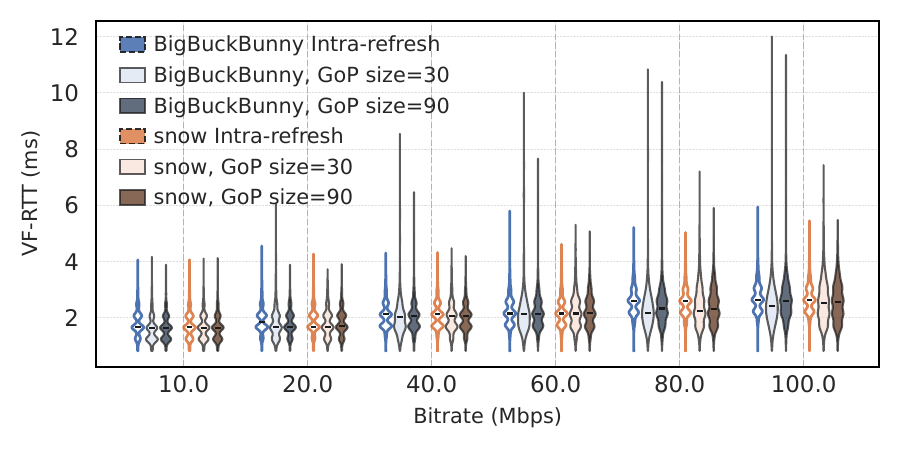}
    \caption{Distribution of frame sizes (top), VMAF scores (middle), and VF-RTT (bottom) for 6 bitrates, with different \gls{gop} and \gls{ir} choices for two video samples streamed at 90 FPS. }
    \label{fig:vmaf_gop}
\end{figure}
% ------------------------------
% ------------------------------
\subsection{Effect of GoP size, \gls{ir}, and bitrate on video quality and latency}

Video frame size distributions vary with the encoded scene and codec settings. We consider a scenario with a single \gls{hmd} placed 1.5~m from the \gls{ap} to isolate the impact of codec parameters such as \gls{gop} size, bitrate, and \gls{ir} on perceptual quality and latency. Fig.~\ref{fig:vmaf_gop} presents the frame size, VMAF, and VF-RTT distributions (using violin plots) for various codec settings and two video samples streamed at 90 FPS. In the plot, each element represents the total range of the observed values, with a black line indicating the median.

Results show \gls{ir} yields significantly less variable frame sizes and VF-RTT than \gls{gop}, though at the cost of slightly lower VMAF due to increased redundancy~\cite{CHEN2015294, Intrarefresh_BER_HEVC}. Larger \gls{gop} sizes reduce variability by decreasing I-frame frequency, improving compression efficiency and slightly increasing VMAF for a given bitrate.
VMAF offers diminishing gains at higher bitrates; a 60~Mbps bitrate achieves good perceptual quality (VMAF $\approx 90$) across all codec settings. While higher bitrates reduce variability on VMAF scores, frame size and latency variability increase with bitrate. In particular, an \gls{ir} stream at 80~Mbps exhibits VF-RTT variance in the range of a 40~Mbps \gls{gop} stream, with similar VMAF ranges. Therefore, our recommendation is prioritizing \gls{ir} with higher bitrates over \gls{gop} at lower bitrates for improved latency stability.

%For higher bitrates, VMAF scores display diminishing returns. Considering that a score of 90 represents excellent perceptual video quality, choosing a bitrate of 60~Mbps already shows decent performance for all \gls{ir} and \gls{gop} choices. Note that, for all choices, the spread of VMAF scores is generally reduced by choosing a higher bitrate, while the spread of frame sizes and latency grows consistently with bitrate. 

% ------------------------------
% ------------------------------
\subsection{Effect of an increasing number of \gls{vr} users}

Next, we analyze network \gls{qos} as the number of simultaneous \gls{vr} users increases, comparing \gls{ir} and \gls{gop} of size 90. Each user streams randomized offsets of the \textit{snow} video sample at 60 or 90 FPS with \gls{cbr} 100 Mbps, all located 1.5~m from the \gls{ap}.
% In this section, we analyze how increasing the number of \gls{vr} users affects \gls{qos} in the network, testing \gls{ir} against a \gls{gop} of size 90. 
% The exact scene streamed to each user is randomized by selecting a random offset for the first $T_\text{chunk}$ in the \textit{snow} video sample. We show results for 60 and 90~FPS. All users are assumed to be at a distance of 1.5~m from the \gls{ap}. 
\begin{figure}[t!!!]
    \centering
     \begin{subfigure}{\columnwidth}
            \centering
            \includegraphics[width=0.99\linewidth]{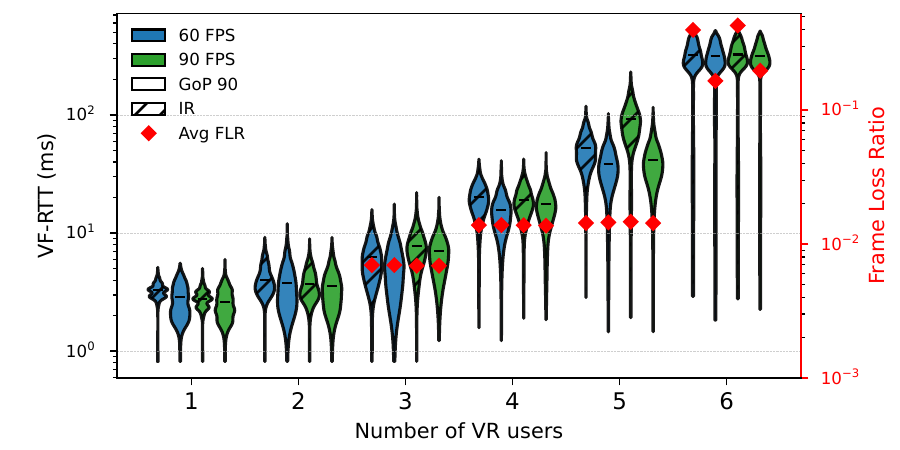}
            \caption{}
            \label{subfig:increasing_users_gop_ir_a}
    \end{subfigure}
    \vspace{-2mm}
     \begin{subfigure}{\columnwidth}
            \centering
            \includegraphics[width=0.99\linewidth]{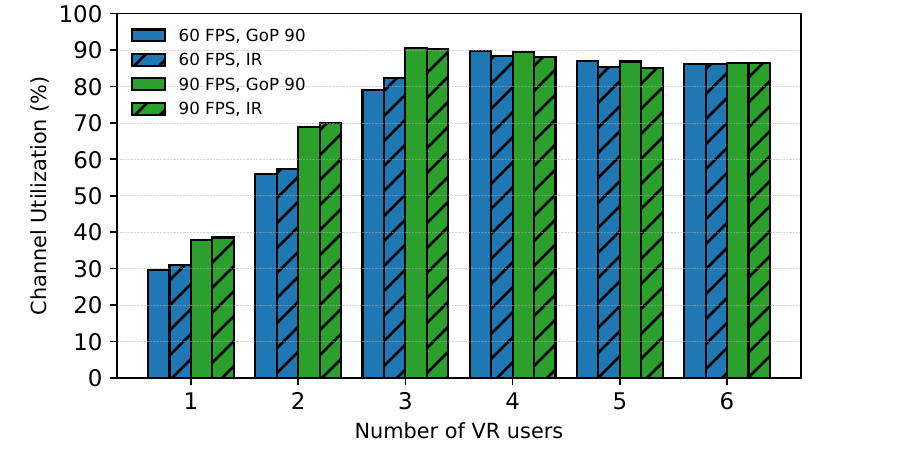}
            \caption{}
            \label{subfig:increasing_users_gop_ir_b}
    \end{subfigure}
        
    \caption{Network performance per number of users, shown through
    % with FPS of 60 or 90 at 100~Mbps, and choice of \gls{ir} vs \gls{gop} of size 90. 
    (a) VF-RTT distribution, FLR average, (b) CU.}
    
    \label{fig:increasing_users_gop_ir}
\end{figure}
Fig.~\ref{fig:increasing_users_gop_ir} shows the distribution of VF-RTT (for correctly received frames) and \gls{flr}, and the mean \gls{cu} as users increase from 1 to 6. The \gls{flr} remains below 1\% up to 3 users; at 4 and 5 users, VF-RTT variability rises sharply (median around 20 and 50 ms), and \gls{flr} exceeds 1\%. The 4-user limit thus marks the threshold for maintaining marginal \gls{qos} with 100~Mbps \gls{cbr} streams in our model.

While \gls{ir} consistently produces lower VF-RTT variance, channel saturation occurs regardless of codec or FPS at 4 users. Higher FPS increases \gls{cu} due to more frequent transmissions (Fig.~\ref{subfig:increasing_users_gop_ir_b}). \gls{cu} starts decreasing at 5-6 users due to larger---and more airtime efficient---\gls{ampdu} transmissions, as a consequence of existing network congestion preventing timely transmission opportunities.

%meaning that client \gls{ul} transmissions cannot reliably access the channel before the next \gls{ul} transmission time. 

% ------------------------------
\captionsetup[subfigure]{skip=5pt} % tighten space under each subcaption
\begin{figure}[t!!]   % star = span both columns
  \centering

  % -------- left-hand figure (was Fig. 1) ----------
  \begin{subfigure}[t]{0.49\textwidth}
    \centering
    % --- first panel ---
    % \begin{subfigure}[b]{\linewidth}
    %   \centering
    %   % \includegraphics[width=0.8\linewidth]{Figures/varying_2xr_1m_violin_peak_network_throughput_bps.pdf}
    %   % \caption{Channel utilization (\%)}
    % \end{subfigure}\\[-0.4ex]
    % --- second panel ---
    \begin{subfigure}[t]{\linewidth}
      \centering
      \includegraphics[width=0.99\linewidth]{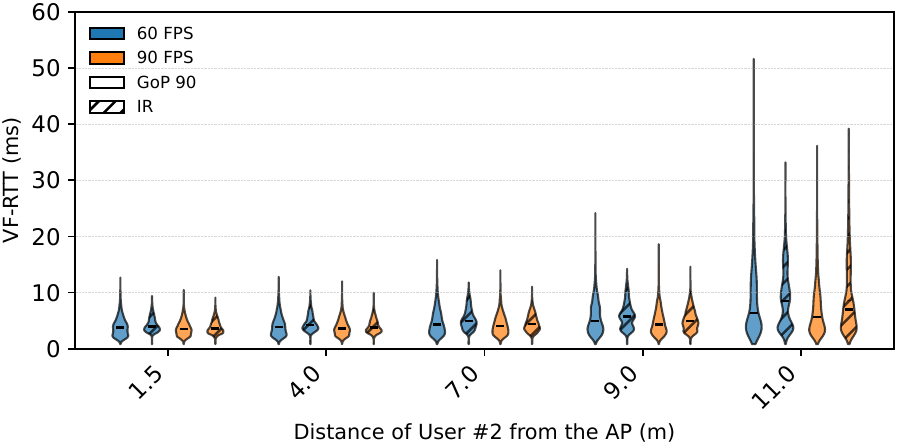}
      % \caption{VF-RTT (ms)}
    \end{subfigure}
    \caption{}
    \label{fig:all-mcs2u_1m}
  \end{subfigure}
  \hfill
  % -------- right-hand figure (was Fig. 2) ----------
  \begin{subfigure}[b]{0.49\textwidth}
    \centering
    % --- first panel ---
    % \begin{subfigure}[b]{\linewidth}
    %   \centering
    %   \includegraphics[width=0.8\linewidth]{Figures/varying_2xr_11m_violin_peak_network_throughput_bps.pdf}
    %   % \caption{Channel utilization (\%)}
    % \end{subfigure}\\[-0.4ex]
    % --- second panel ---
    \begin{subfigure}[b]{\linewidth}
      \centering
      \includegraphics[width=0.99\linewidth]{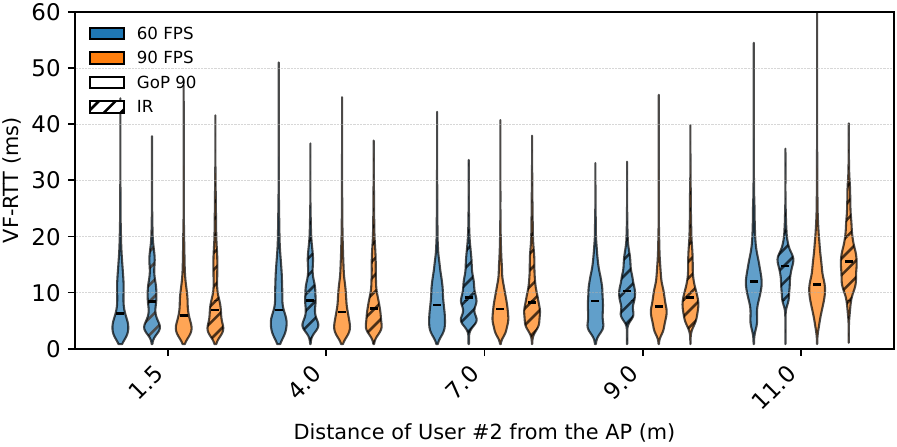}
      % \caption{VF-RTT (ms)}
    \end{subfigure}
    \caption{}
    \label{fig:all-mcs2u_11m}
  \end{subfigure}
  \caption{Distribution of VF-RTT of two users, one placed at a fixed location and a second one moving away from the \gls{ap}. 
  % 60/90 FPS, \gls{ir} vs \gls{gop} of size 90 are compared.
  (a) User~\#1 is at 1.5~m from the AP, (b) User~\#1 is at 11~m from the AP.}
  \label{fig:distance-comparison}
\end{figure}

% ------------------------------
% ------------------------------
\subsection{Effect of \gls{ap}-\gls{hmd} distance and heterogeneous \gls{mcs}}

Transmission rates depend on \gls{mcs}, which varies with user distance. Users with lower \gls{mcs} consume more channel airtime, significantly increasing queuing delays for others—a phenomenon known as the Wi-Fi performance anomaly~\cite{bellalta2014throughput}. We evaluate two \gls{vr} users streaming the \textit{snow} video at 100~Mbps and 60/90 FPS, fixing the position of one user (User \#1) at either 1.5~m (Fig.~\ref{fig:all-mcs2u_1m}) or 11~m (Fig.~\ref{fig:all-mcs2u_11m}) from the \gls{ap}, and varying the distance of the other user (User \#2) to induce \gls{mcs} differences. %The other user is fixed at either 1.5~m (Fig.~\ref{fig:all-mcs2u_1m}) or 11~m (Fig.~\ref{fig:all-mcs2u_11m}) from the \gls{ap}. 
Both \gls{ir} and \gls{gop} of size 90 are considered. Fig.~\ref{fig:distance-comparison} confirms that as one \gls{hmd} (User \#2) moves further from the \gls{ap}, latency \gls{qos} degrades not only for that user but also for the other at a favorable location (User \#1). In Fig.~\ref{fig:all-mcs2u_1m}, the VF-RTT increase for User \#1 arises from queuing delays induced by User \#2's lower \gls{mcs} transmissions. This delay stems from the time taken to serve all downlink traffic to the distant \gls{sta} or uplink transmissions from it. The median and variance of VF-RTT grow with the User \#2's separation, independently of codec or FPS choices. When User \#1 is at 11~m from the \gls{ap} (Fig.~\ref{fig:all-mcs2u_11m}), latency degradation is more pronounced, exceeding the 33~ms threshold across all configurations.

% % ------------------------------
% % ------------------------------
% % ------------------------------
% ------------------------------
\section{Conclusions}

In this paper, we presented a \gls{vr} streaming emulation framework for the joint evaluation of Wi-Fi scenarios and video codec configurations. The simulation results illustrate how network performance varies with different codec and network parameter choices, and we provide recommendations based on these observations. Specifically, the results highlight the trade-off between video quality through VMAF, and frame size for \gls{gop} versus \gls{ir}-based  coding strategies. 
% \gls{ir} proves preferable due to its more consistent frame size distribution.
The channel saturates with more than four concurrent users independently of the GoP or \gls{ir} choice. For a lower number of users (before saturation) \gls{ir} exhibits reduced variance in VF-RTT.   
% The channel still saturates when more than four concurrent users are present.
In future work, we plan to expand the proposed framework into an open testbed to evaluate \gls{abr} algorithms and advanced Wi-Fi solutions. This includes support for \gls{dl}/\gls{ul} coordinated \gls{xr} traffic scheduling, Multi-Link Operation
% ~\cite{carrascosa2024performance}
and Multi-AP coordination 

% ~\cite{nunez2025improving}.

% % ------------------------------
% % ------------------------------
% % ------------------------------
% ------------------------------
% \section{Acknowledgement}

% This work is partially funded by MAX-R (101070072) EU, Wi-XR PID2021-123995NB-I00 (MCIU/AEI/FEDER,UE), and by MCIN/AEI under the Maria de Maeztu Units of Excellence Programme (CEX2021-001195-M). %\bb{Ferran Maura is also supported by FPI XXXXX}. 

% ------------------------------
% ------------------------------
% ------------------------------
% ------------------------------
% 0.5 pages for references (10-15 max).
%\bibliographystyle{plain}
%\bibliographystyle{elsarticle-num} 
\bibliographystyle{IEEEtran} 
\bibliography{bib}

\end{document}